\documentclass[twocolumn,prb,aps,floatfix,superscriptaddress]{revtex4-1}

\usepackage{color}
\usepackage{siunitx}

\usepackage{graphicx}
\usepackage[caption=false]{subfig}

\usepackage{amsthm}
\usepackage{amsmath}
\usepackage{amssymb}
\usepackage{enumerate}

\DeclareMathOperator{\Tr}{Tr}

\newcommand{\ket}[1]{\vert #1 \rangle}
\newcommand{\eket}[1]{\bigl \vert #1 \bigr \rangle}
\newcommand{\R}{\boldsymbol{R}}
\newcommand{\lo}{\ell_0}
\newcommand{\li}{\ell_{\rm int}}

\newcommand{\ebra}[1]{\bigl \langle #1 \bigr \vert}
\newcommand{\eexp}[1]{\bigl \langle #1 \bigr \rangle}
\newcommand{\figref}[1]{Fig.~\ref{#1}}

\newcommand{\ren}{R\'{e}nyi~}

\begin{document}

\title{Particle partition entanglement of bosonic Luttinger liquids}

\author{C. M. Herdman}
\email{cherdman@uwaterloo.ca}
\altaffiliation[Present Address: ]{Institute for Quantum Computing, University of Waterloo, Ontario, N2L 3G1, Canada}
\affiliation{Institute for Quantum Computing, University of Waterloo, Ontario, N2L 3G1, Canada}
\affiliation{Department of Physics \& Astronomy, University of Waterloo, Ontario, N2L 3G1, Canada}
\affiliation{Department of Chemistry,  University of Waterloo, Ontario, N2L 3G1, Canada}
\affiliation{Department of Physics, University of Vermont, Burlington, VT 05405, USA}

\author{A. Del Maestro}
\affiliation{Department of Physics, University of Vermont, Burlington, VT 05405, USA}
\affiliation{Vermont Complex Systems Center, University of Vermont, Burlington, VT 05405, USA}

\begin{abstract}
We consider the R\'{e}nyi entanglement entropy of bosonic Luttinger liquids under a particle bipartition and demonstrate that the leading order finite-size scaling is logarithmic in the system size with a prefactor equal to the inverse Luttinger parameter. While higher order corrections involve a microscopic length scale, the leading order scaling depends only on this sole dimensionless parameter which characterizes the low energy quantum hydrodynamics. This result contrasts the leading entanglement entropy scaling under a spatial bipartition, for which the coefficient is universal and independent of the Luttinger parameter. Using quantum Monte Carlo calculations, we explicitly confirm the scaling predictions of Luttinger liquid theory for the Lieb-Liniger model of $\delta$-function interacting bosons in the one dimensional spatial continuum.
\end{abstract}

\maketitle

\section{Introduction}

Entanglement is a fundamental non-classical property of quantum many-body systems, and provides a physical resource for quantum information processing~\cite{Horodecki2009}. The quantification of entanglement via entanglement entropy (EE) has been demonstrated to aid in the characterization of quantum matter \cite{Amico2008,Eisert2010}, especially when the conventional use of local correlation functions may not be feasible due to a lack of symmetry breaking. In particular, the study of the EE between two spatially distinct regions (partitions) of a many-body system allows the canonical ``area law" scaling to be probed, including its sub-leading corrections which have been demonstrated to reveal fundamental features of the underlying phase of matter~\cite{Srednicki1993, Hamma2005c,Hamma2005d,Levin2006a, Kitaev2006b,Fradkin2006,Casini2007,Metlitski2009,Stephan2009, Hsu2010,Zaletel2011,Metlitski2011,Kallin2011,Ju2012,Stephan2012}.  Systems of itinerant particles can alternatively be bipartitioned into subsets of \emph{particles} rather than spatial regions~\cite{Zanardi2002, Shi2003,Fang2003,Zozulya2007a,Haque2007a,Zozulya2008,Haque2009},  with the associated entanglement entropies providing complementary information on the role that interactions and exchange statistics play in generating the non-classical correlations in a quantum state. 

The study of EE has been particularly fruitful in low dimensional quantum many-body systems, where strong fluctuations inhibit transitions to conventional ordered phases.  For example, the EE of a spatial bipartition of a critical system in one-dimension (1D) scales logarithmically with subsystem size, where the prefactor is proportional to the central charge of the underlying conformal field theory~\cite{Vidal2003,Calabrese2004}. The low energy behavior of such systems is known to be described by Luttinger liquid (LL) theory, which characterizes the universal long distance physics that is unique to quantum mechanics in one spatial dimension~\cite{Haldane1981}.  A LL is characterized by a single non-universal dimensionless parameter $K$ which determines the nature of correlations functions, in addition to a velocity which sets the energy scale for quasiparticle excitations. While the central charge is a universal feature of LLs, $K$ depends on the specific details of the system, and in general, is not easily determined for a given microscopic model. Thus the leading spatial entanglement scaling gives access to a universal quantity of the LL, but is not sensitive to $K$ which characterizes the long-distance behavior of correlation functions. The Luttinger parameter does appear in the bipartite fluctuations of a Luttinger liquid, which scales logarithmically with subsystem size with a prefactor that is proportional to $K$~\cite{Song2010,Song2011}, as well as higher order corrections to the spatial partition EE~\cite{Laflorencie2006} and the EE of a partition into disjoint spatial regions~\cite{Furukawa2009,Calabrese2009b}.

In this paper we demonstrate that the finite-size scaling form of the $n$-\emph{particle partition} 2nd \ren entanglement entropy of a $N$-particle bosonic Luttinger liquid is:  
\begin{equation}
    S_2 \left( n \right) \simeq \frac{n}{K} \log N +\mathrm{const.}  + 
    \mathcal{O}\left(\frac{1}{N^{1-1/K}}\right);  
    \label{eq:S2LL}
\end{equation}
thus the leadinger-order scaling coefficient of the particle partition EE is proportional to the the inverse Luttinger parameter. We evaluate the explicit form of the constant and correction terms for a bosonic LL, and confirm their accuracy for the Lieb-Liniger model of $\delta$-function interacting bosons in 1D~\cite{Lieb1963,Lieb1963a} via large scale quantum Monte Carlo simulations.
%
\begin{figure}[t]
\begin{center}
\includegraphics[width=1.0\columnwidth]{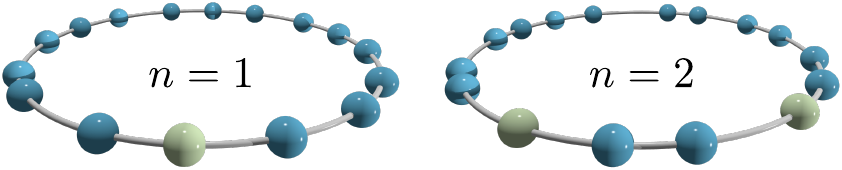}
\end{center}
\caption{Systems of $N$ itinerant bosons in the one-dimensional spatial continuum under $n$-particle partitions indicated through a fictitious coloring scheme.}
\label{fig:partition}
\end{figure}
%

\section{Quantum hydrodynamics}

The ubiquity of Luttinger liquid behavior manifests in its experimental observation in a diverse set of quasi-1D experimental systems including electronic conductance and tunneling in carbon nanotubes \cite{Bockrath:1999cd} and quantum Hall edges \cite{Chang:2003cl} as well as in the energy and correlation functions of low density ultra cold bosonic gases \cite{Paredes:2004fp, Kinoshita:2004jpa, Kinoshita2005}. Recent work has proposed that it may also be seen in confined superfluids of ${}^4$He \cite{DelMaestro2011,Eggel:2011fj} with associated experiments already underway \cite{Taniguchi:2010dh}.

Here, we focus on spinless bosons, and the low energy effective Hamiltonian capturing the hydrodynamics of a LL with $N$ interacting particles on a line of length $L$  with density $\rho_0 = N/L$ may be written in terms of bosonic fields $\phi(x)$ and $\theta(x)$ representing the phase and density fluctuations of a microscopic field operator 
$\psi^{\dagger}$ as: 
\begin{equation} 
H = \frac{\hbar v}{2 \pi} \int dx~\left[ K\left(\frac{\partial \phi}{\partial x}\right)^2 +\frac{1}{K}\left(\frac{\partial \theta}{\partial x}\right)^2 \right] 
\label{eq:H_LL}
\end{equation}
where
\begin{align}
\psi^{\dagger}(x) \simeq \sqrt{\rho_0 + \frac{1}{\pi}\frac{\partial\theta}{\partial x}} e^{-i\phi(x)}, \\
\left[\phi(x),\frac{\partial\theta(x')}{\partial x'} \right] = i \pi \delta(x-x').
\end{align} 

For Galilean invariant systems, the Luttinger parameter $K$ may be computed from the compressibility, as
\begin{equation}
    K = \frac{\hbar \pi}{L} \left(\frac{m}{N}\frac{\partial^2 E_0}{\partial N^2}\right)^{-1/2}, 
\end{equation}
where $E_0$ is the ground state energy and $m$ is the mass of the particles.  Even at zero temperature, LLs display no true long-range order; the ground state of Eq.~\eqref{eq:H_LL} has algebraically decaying correlations parametrized by $K$. In the non-interacting limit, $K\rightarrow\infty$, and phase correlations dominate over density correlations. The reverse is true for impenetrable bosons with $K \to 1^{+}$.  In particular, the one-body reduced density matrix is a power law at long distances, with exponent proportional to $K^{-1}$ :
\begin{equation}
\rho_1\left(x,x'\right) \equiv \left \langle \psi^\dagger(x)\psi(x') \right \rangle \sim \left \vert x-x'\right\vert^{-1/2K}.
\label{eq:rho1}
\end{equation}
With the exception of a few models where exact solutions are available \cite{Girardeau:1960ff, Lieb1963, Sutherland:1971dp}, determining the precise value of $K$ for a given microscopic model is a difficult task, and may require a detailed analysis of the decay of correlation functions \cite{Astrakharchik:2003dc}.  We find that this information is encoded in the leading order scaling of the particle partitioned entanglement entropy, and can even be bounded by measurements of the condensate fraction \cite{Herdman2014}.

 \section{Particle partition entanglement}

Given a bipartition of a system into subsystems $A$ and $B$, one can quantify the entanglement in $\ket{\Psi_{AB}}$ describing the full pure state via \ren entanglement entropies:
\begin{equation}
S_{\alpha}\left[ \rho_A \right] \equiv \frac{1}{1-\alpha} \log \left({ {\rm Tr} \rho_A^{\alpha} }\right), \label{eq:renyi}
\end{equation}
where $\rho_A$ is the reduced density matrix (RDM) of subsystem $A$,
\begin{equation}
\rho_A =  {\rm Tr}_B \biggl( \eket{\Psi_{AB}}\ebra{\Psi_{AB}}\biggr) \notag
\end{equation}
and $\alpha$ is the \ren index. For $\alpha \rightarrow 1$ the R\'enyi entropy is equivalent to the von Neumann entropy: $-\mathrm{Tr}\, \rho_A \log \rho_A$. 

Entanglement entropy of many-body systems is most often studied under spatial bipartitions. For example, if a translationally invariant 1D system is bipartitioned into contiguous subsystems, then subsystem $A$ is completely characterized by the length of the subsystem $\ell_A$. For the ground states of critical 1D systems, the leading order scaling of the EE is logarithmic in $\ell_A$~\cite{Vidal2003,Calabrese2004},
 \begin{equation}
S\left(\ell_A\right) \approx \frac{c}{3} \log \ell_A +\dots,
\end{equation}
where $c$ is the central charge of the underlying conformal field theory. 
In systems of itinerant particles, one may alternatively bipartition the system into subsets of particles (Fig.~\ref{fig:partition}), rather than spatial subregions~\cite{Zanardi2002, Shi2003,Fang2003,Zozulya2007a,Haque2007a,Zozulya2008,Haque2009}. An indistinguishable particle partition is uniquely characterized by the number of particles $n$ in the subsystem, with the RDM for the subsystem given by the $n$-body reduced density matrix:
\begin{equation*}
\rho_n  \equiv \int \! d\boldsymbol{r}_{n}\dots \int\! d\boldsymbol{r}_{N-1}
 \left \langle \boldsymbol{r}_{n}\dots \boldsymbol{r}_{N-1} \right \vert \rho_N \left \vert \boldsymbol{r}_{n}\dots \boldsymbol{r}_{N-1} \right \rangle, \notag
\end{equation*}
where $\{\boldsymbol{r}_i\}$ are the positions of the particles and $\rho_N$ is the full $N$-body density matrix. One can then quantify the entanglement entropies between subsets of particles by the particle partitioned EE~\cite{Zozulya2008,Haque2009}:
\begin{equation}
S_\alpha \left(n\right) \equiv S_\alpha \left[ \rho_n \right].
\end{equation}
Note that since $\rho_n$ is non-local in space and contains no length scale, $S_\alpha (n)$ can capture physics that is distinct from, and complimentary to, the EE under a spatial bipartition.

The importance of particle partition EE, and in particular, its connection to physically utilizable entanglement for quantum information processing applications has only recently been uncovered \cite{Wiseman:2003jx, Killoran:2014gu}.  
Zozulya \emph{et al.} have proposed a general scaling form for particle EE which is linear in the subsystem size and logarithmic in the system size \cite{Zozulya2008}:
\begin{equation}
S\left(n,N\right) = b_1 n\log N + b_2 .
\end{equation}
The coefficients $b_1,b_2$ have only been computed in a limited number of models and one may ask what physical content is contained within\cite{Santachiara2007,Zozulya2007a,Haque2007a,Zozulya2008,Haque2009}.  Additionally, previous work has demonstrated that in systems of itinerant bosons, the $n=1$ second \ren EE is bounded by the experimentally accessible condensate fraction, thus offering a direct relationship between particle EE and experimentally measurable quantities~\cite{Herdman2014}.

 \section{One particle entanglement in bosonic Luttinger Liquids}

\subsection{Asymptotic scaling for $L\rightarrow \infty$}

We begin with an analysis of the one particle partition 2nd \ren EE $S_2(n=1)$ for a LL described by Eq.~\eqref{eq:H_LL}.  Using the Galilean invariance of the one-body RDM,
we take the general, appropriately normalized $(\Tr \rho_1 = 1)$ and regularized form of $\rho_1$ to be:
\begin{align}
    \rho_1 (r) = \frac{1}{L}\begin{cases}
    \Phi\left(\frac{r}{a}\right) & r \leq a \\
\Phi(1) \left(\frac{a}{r}\right)^{1/2K} & r > a
\end{cases} \label{rho1Form}
\end{align}
where we have used Eq.~\eqref{eq:rho1}, $a$ is some short distance length scale and $\Phi(y)$ is a dimensionless function satisfying $\Phi(y) \leq 1$, $\Phi(0) = 1$ that depends on the non-universal details of the microscopic model.  Performing the trace in Eq.~\eqref{eq:renyi} (see Refs.~\onlinecite{Gritsev2006,Polkovnikov2006} for related calculations) for $\alpha=2$ yields
\begin{align*}
\Tr \rho_1^2 &= \int_{-L/2}^{L/2} dx dx' \rho_1\left(x,x'\right)\rho_1\left(x',x\right) \notag \\
&= \Phi\left(1\right)^2 \frac{K}{K-1} \left(\frac{2a}{L}\right)^{1/K} \notag \\
&\quad\quad \times \Biggl \{ 1 +\left(\frac{2a}{L}\right)^{1-1/K}\left[\frac{\overline{\Phi^2}(1)}{\Phi(1)^2}\frac{K-1}{K}-1\right] \Biggr \}
\end{align*}
where 
\begin{equation}
\overline{\Phi^2}(y) \equiv y^{-1} \int_0^{y} \Phi(z)^2 dz.
\label{eq:defPhi2}
\end{equation}  
Taking the log, the central finding of our work is that the finite size scaling of the one-particle 2nd \ren EE for $K>1$ is described by 
\begin{align}
S_2(n=1) &\simeq b_1\log N + b_2 - \log\left[1+\frac{b_3}{N^{1-b_1}} \right]
\label{eq:S2NFS}
\end{align}
where $b_1 = 1/K$ and the constants $b_{2,3}$ depend on both $K$ as well as the short distance physics captured in $\Phi(r/a)$.  We note this result is consistent with previously reported results for the Tonks-Girardeau model of impenetrable bosons, which corresponds to $K\rightarrow1$ where it was found that $S_1(n=1) \sim \ln N$~\cite{Santachiara2007,Haque2009}. Thus the leading scaling coefficient of the particle EE is equal to the inverse of the Luttinger parameter which characterizes the universal quantum hydrodynamics in one dimension.

\subsection{Finite size corrections}

Modifications  can be determined for a finite system with periodic boundary conditions by replacing the separation $|x-x'|$ in Eq.~\eqref{eq:rho1} with 
the distance between two points on a ring of circumference $L$ ~\cite{Cazalilla2004,Cazalilla2011}:
\begin{equation}
    \left \vert x-x' \right \vert \rightarrow \frac{L}{\pi}\sin\left(\frac{\pi} {L}\left|x-x'\right|\right).
\end{equation}
Proceeding as above, by assuming a short distance form $\Phi(r/a)$ and by employing normalization and translational invariance we may write:
\begin{equation}
    \rho_1(r|L) = \frac{1}{L}\begin{cases}
            \Phi(r) & r \le a\\
            \Phi(1) \left \lvert \frac{\sin(\pi a/L)}{ \sin(\pi r /L)} \right \rvert^{1/2K}
            & r > a
        \end{cases}. 
\end{equation}
Performing the trace over $\rho_1^2$ now yields:
\begin{widetext}
\begin{align}
    \mathrm{Tr}\, \rho_1^2 
                           &= \frac{2a}{L} \left[ \overline{\Phi^2}(1) + \Phi(1)^2\left|\sin \frac{\pi a}{L}\right|^{1/K}
\int_1^{L/2a} d \overline{r}\csc^{1/K}\left(\frac{\pi a \overline{r}}{L}\right)\right] \nonumber \\
&= \frac{2a}{L} \left[\overline{\Phi^2}(1) + \Phi(1)^2\frac{L}{\pi a}\left \lvert \sin \frac{\pi a}{L}\right|^{1/K}
\cos \frac{\pi a}{L}\; {}_2F_1\left(\frac{1}{2},\frac{1}{2} + \frac{1}{2K};
\frac{3}{2}; \cos^2 \frac{\pi a}{L} \right)\right] 
\label{seq:Trrho12}
\end{align}
where ${}_2F_1(q,b;c;z)$ is the Hypergeometric function.  Thus we find the \ren entropy to be
\begin{align}
\!\!  S_2(n=1) &= -\frac{1}{K}\log \sin\! \frac{\pi \rho_0 a }{N} - \log \left [ 
    \frac{2 \rho_0 a \overline{\Phi^2}(1)}{N \sin^{1/K}(\pi \rho_0 a/N)}
    + \frac{2}{\pi}\Phi(1)^2 \cos \frac{\pi \rho_0 a}{N}\;
{}_2F_1\left(\frac{1}{2},\frac{K+1}{2K}; 
\frac{3}{2}; \cos^2 \frac{\pi \rho_0 a}{N} \right)\right]. \label{eq:S2NFS1}
\end{align}
Expanding for large $N$ and making use of the $\Gamma$-function recursion relation:
$\Gamma(z)\Gamma(1-z) = -z \Gamma(z)\Gamma(-z) = \pi \csc \pi z$ we find: 
\begin{align}
S_2(n=1) &\simeq \frac{1}{K}\log N - \log \left \{ 
    1 + \frac{2\left[(K-1)\overline{\Phi^2}(1) - K \Phi(1)^2 \right]\pi^{{3}/{2}-{1}/{K}}}{(K-1)\Phi(1)^2 
\sin \frac{\pi}{2K} \Gamma\left(\frac{1}{2K}\right) \Gamma\left(\frac{1}{2}-\frac{1}{2K}\right)} \left(\frac{\rho_0 a}{N}\right)^{1-\frac{1}{K}} \right \} \nonumber \\
& \qquad\qquad\qquad \qquad- \log \left[ 
\pi^{{1}/{K}-{3}/{2}}(\rho_0 a)^{1/K}
\Phi(1)^2\sin \frac{\pi}{2K} \Gamma\left(\frac{1}{2K}\right) \Gamma\left(\frac{1}{2}-\frac{1}{2K}\right)
\right],
\label{seq:S2N}
\end{align}
\end{widetext}
which exhibits identical scaling behavior as that in Eq.~\eqref{eq:S2LL} and \eqref{eq:S2NFS}, albeit with different non-universal subleading coefficients $b_{2,3}$. Up to this point, we have only demonstrated the accuracy of Eq.~\eqref{eq:S2LL} for $n=1$, but we now conjecture its leading order $n$ dependence for a Luttinger liquid under an arbitrary particle partition of size $n$, and numerically test for $n\ge1$ in a specific microscopic model. 

\section{Confirmation in the Lieb-Linger model}
The Lieb-Liniger model describes $N$ non-relativistic bosons interacting with a contact potential in 1D~\cite{Lieb1963a,Lieb1963}, with Hamiltonian:
\begin{equation}
    H = -\lambda \sum_{i=1}^N \frac{d^2}{d x_i^2} + g\sum_{i<j} \delta\left(x_i-x_j\right), 
    \label{eq:HLieb}
\end{equation}
where $\lambda \equiv \hbar^2/2m$ and $g$ is an interaction strength with dimensions of energy $\times$ length. We consider only repulsive interactions $g\geq0$ and as $g\rightarrow + \infty$ the Tonks-Girardeau \cite{Girardeau:1960ff} gas of impenetrable bosons is recovered. It is useful to parameterize finite interactions using a single dimensionless parameter $\gamma \equiv gL/2\lambda N$, which has particular experimental relevance to ultracold Bose gases confined in quasi-1D optical traps.\cite{Kinoshita2005}  
The Lieb-Liniger model is analytically soluble via the Bethe ansatz, which provides access to the exact ground state energy in the thermodynamic limit $L\to\infty$. \cite{Lieb1963} At low energies, its emergent properties are well described by LL theory and moreover, the existence of an exact solution allows for access to the short distance properties of the one-body RDM parameterized here by the function $\Phi(r/a)$ \cite{Olshanii2003}. 

\begin{figure}[t]
\begin{center}
\includegraphics[width=0.95\columnwidth]{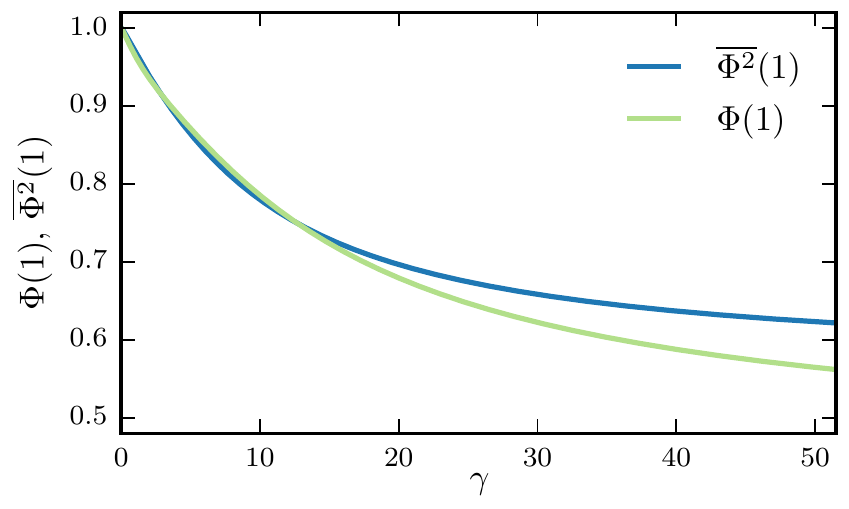}
\end{center}
\caption{Short distance properties of the one body reduced density matrix as a function of the dimensionless interaction parameter $\gamma = g/2\lambda\rho_0$ for the Lieb-Liniger model. $\Phi(1)$ and $\overline{\Phi^2}(1)$ are the bare and integrated short distance functions given by Eq.~\eqref{eq:LLfr} at the crossover microscopic length scale $r = a$ appearing in the particle partition entanglement entropy.}
\label{fig:Phi}
\end{figure}

Olshanii and Dunjko \cite{Olshanii2003} have analyzed the short distance behavior of the one-body density matrix $\Phi(r/a)$ as a power series
\begin{equation}
    \Phi\left(\frac{r}{a}\right) = 1 + c_2 \left(\frac{r}{a}\right)^2 + c_3 \left(\frac{r}{a}\right)^3 + \cdots
    \label{eq:LLfr}
\end{equation}
where $r \le a$ and the constants $c_2$ and $c_3$ were found to be:
\begin{align}
\label{eq:c23}
c_2 = -\frac{1}{2} \left[e(\gamma) - \gamma e'(\gamma) \right],\qquad c_3 = \frac{1}{12} \gamma^2 e'(\gamma)
\end{align}
with $e(\gamma)$ a dimensionless function of the interaction parameter $\gamma =  {g}/{2\lambda \rho_0}$ appearing in the ground state energy $E_0 = \lambda \rho_0^2 N e(\gamma)$ that can be numerically determined from the Bethe ansatz solution \cite{Lieb1963}. 
Both $\Phi(1)$ and $\overline{\Phi^2}(1)$ are decaying functions of the interaction parameter $\gamma$ as shown in Fig.~\ref{fig:Phi}.
%
%

 \subsection{Quantum Monte Carlo Method}

We test our LL calculations for the particle partition EE of the Lieb-Liniger model using a path integral ground state quantum Monte Carlo (QMC) method \cite{Ceperley1995,Sarsa2000} which provides unbiased access to ground state expectation values through imaginary time projection: 
\begin{equation}
\eexp{ \hat{\mathcal{O}} } = \lim_{\beta \rightarrow \infty}\frac{ \ebra{\Psi_{\mathrm{T}}}e^{-\beta H/2} \hat{\mathcal{O}} e^{-\beta H/2} \eket{\Psi_{\mathrm{T}}}}{\ebra{\Psi_{\mathrm{T}}}e^{- \beta H} \eket{\Psi_{\mathrm{T}}}} \label{eq:O_PIGS}
\end{equation}
where $ \hat{\mathcal{O}}$ an observable and $\ket{\Psi_{\mathrm{T}}}$ is a trial wave function. In this section, for ease of presentation, we set $\hbar=1$. We Monte Carlo sample the discrete imaginary time worldlines of $N$ bosons in one spatial dimension with the $\delta$-function interactions described by  the Lieb-Linger Hamiltonian in Eq.~\eqref{eq:HLieb}. Using a discrete imaginary time representation, we approximate the propagator as the product of short time propagators $e^{-\beta H} \simeq \left( \rho_\tau \right)^P$ where  $P \equiv \beta/\tau$. We represent the imaginary time worldlines configurations in the position basis, such that each imaginary time slice is described by a state $\ket{\R}$, where $\R = \{r_0,\ldots,r_{N-1}\}$ is a vector of length $N$ describing the position of all particles in continuous space. The short time propagator $\rho_\tau$ is approximately decomposed into the product of the free particle propagator, which can be sampled exactly, and an interaction propagator such that
\begin{align}
 \rho_\tau (\R,\R') &= \rho_0(\R,\R';\tau,\lambda) \rho_{\rm{int}}(\R,\R';\tau) \\ 
 &\simeq  \ebra{\R} e^{-\tau H} \eket{\R'},
\end{align}
where $\rho_0(\R,\R';\tau,\lambda)$ is the free $N$ particle propagator
\begin{align}
    \rho_0(\R,\R';\tau,\lambda) &\equiv \prod_{j=0}^{N-1} \rho_0 \left(r_j-r_j',\tau,\lambda \right),\\
    \rho_0 \left(\Delta x,\tau,\lambda\right) &\equiv \frac{e^{-\Delta x^2/4\lambda \tau}}{2 \sqrt{\pi \lambda \tau}} .
\end{align}

To efficiently sample the imaginary time propagator, because of the short-ranged nature of the interactions we use a pair-product decomposition \cite{Ceperley1995} which employs the exact two-body propagator for $\delta$-function interacting bosons~\cite{Gaveau1986,Casula2008,Manoukian1989}:
\begin{equation}
 \rho_{\rm{int}} \left(\R, \R';\tau\right) \simeq \prod_{j \neq k} W_{\rm int} \left(r_j-r_k,r'_j-r'_k;\tau\right)
\end{equation} 
where $W_{\rm int}$ is a weight that takes into account the pairwise interactions, and only depends on the relative separation of each pair. The form of $W_{\rm int}$ for the Lieb-Linger model is given in Appendix~\ref{app:PPLL}.

\ren EE are accessible via QMC by sampling an extended configuration space consisting of two identical replicas of the physical system under consideration. The sampled ensemble has imaginary-time worldlines that are \emph{broken} at the center of both paths corresponding to the ``A" subsystems consisting of $n$ particles \cite{Hastings2010,Herdman2014a,Herdman2014}. The estimator for $\Tr \rho_A^2$ is related to the expectation value of the short-imaginary-time propagator which connects the broken worldlines across the replicas:
\begin{equation}
\Tr \rho_A^2 = \frac{\left\langle \rho_A^{\mathrm{SWAP}} \right \rangle}{\left\langle \rho_A^{\mathrm{DIR}} \right \rangle},
\end{equation}
where $\rho_A^{\mathrm{DIR}}$ and $\rho_A^{\mathrm{SWAP}}$ are the reduced propagators for the A subsystems which connect the broken beads to the same and other replica, respectively; see Ref.~\onlinecite{Herdman2014a} for complete algorithmic details.

 \subsection{Numerical Results}

Using QMC, we have computed $S_2(n)$ for the ground state of the Lieb-Linger Model for $n=1,2$ and $\gamma$ and $N$ ranging over two orders of magnitude. 
%
\begin{figure}[t]
\begin{center}
\scalebox{1}{\includegraphics[width=\columnwidth]{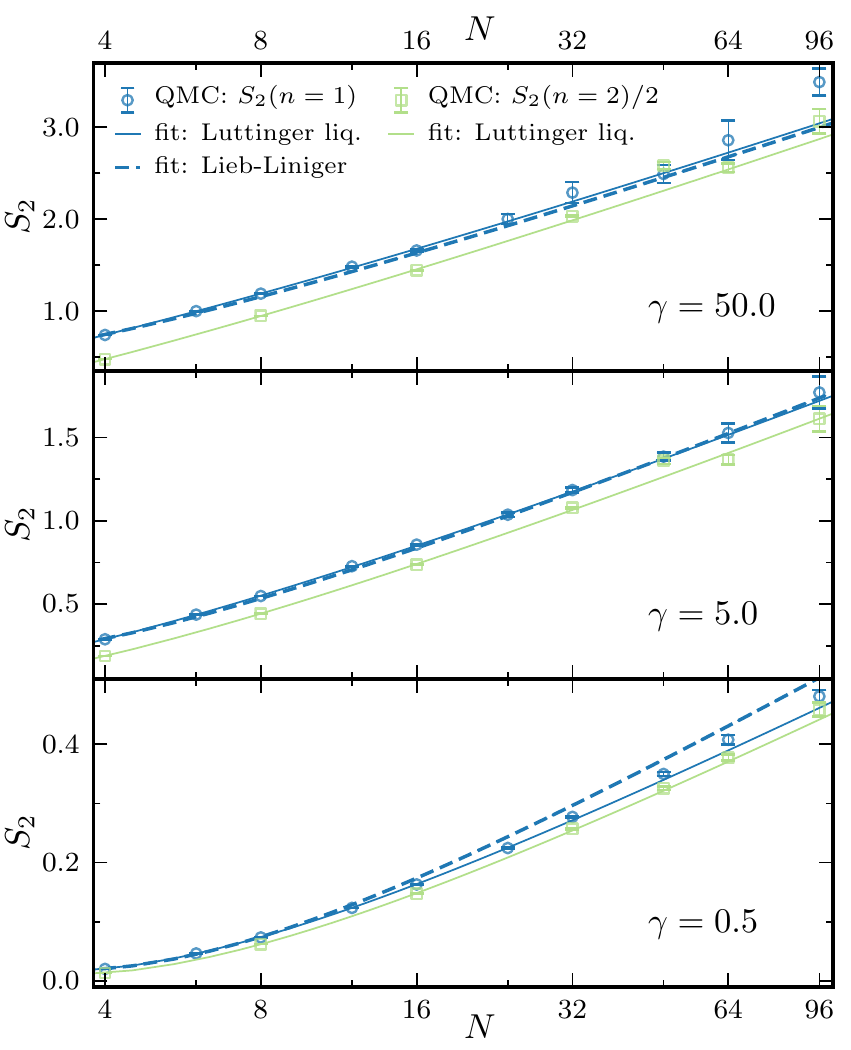}}
\end{center}
\caption{(Color online) Finite-size scaling of $S_2(n)$ of ground state of the Lieb-Linger model for $\gamma=0.5,5,50$ for $4\leq N \leq 96$. The solids lines represent three parameter fits to the generic Luttinger liquid (Eq.~\eqref{eq:S2NFS}) while the dashed line is a one-parameter fit to the finite-size Lieb-Liniger form of the scaling coefficients in Eq.~\eqref{eq:S2NFS1}. $S_2(n=2)$ is shown scaled by a factor of $2$.} 
\label{fig:LLFSS}
 \end{figure}
%
 \figref{fig:LLFSS} shows the finite-size scaling of $S_2(n)$ at constant $\gamma$ for $4\leq N\leq96$ and $\gamma\in\{0.5,5,50\}$. The EE data has been analyzied via two approaches: (solid lines) a generic LL scaling form and (dashed lines) the microscopic finite-size form for the Lieb-Linger model. The generic LL scaling form is given in Eq.~\eqref{eq:S2NFS} where $b_{1-3}$ are taken to be free parameters. The coefficient of the $\log N$ term, $b_1$ allows us to determine $K$ numerically, directly from the QMC data, free of any assumptions regarding the microscopic model. Alternativerly, for each value of $\gamma$, we may use the short distance behavior of the Lieb-Linger one-body RDM \cite{Olshanii2003} and $K(\gamma)$ computed from the Bethe ansatz \cite{Lieb1963a} in combination with Eq.~\eqref{eq:S2NFS1}, leaving only a single free parameter: the short distance cutoff $a$. \figref{fig:avg} shows the best fit values of $a$ as a function of $\gamma$; for all $\gamma$ considered here, we find that $a$ is of order $\rho_0^{-1}$, a natural short-distance length scale.
%
\begin{figure}
\begin{center}
\includegraphics[width=0.95\columnwidth]{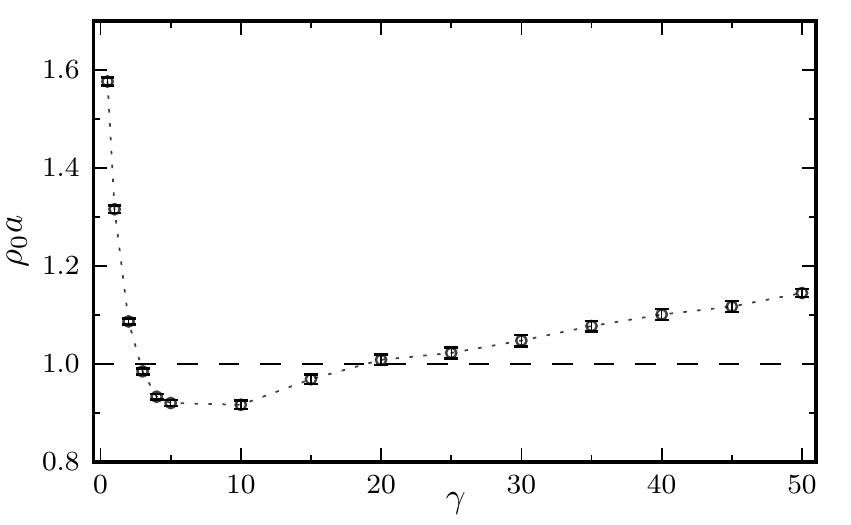}
\end{center}
\caption{ Best fit for the short distance cutoff $a$ for the one-parameter fit to the finite-size Lieb-Linger scaling form to the finite-size scaling of $S_2(n=1)$ as shown in  \figref{fig:LLFSS} and \figref{fig:SvsNb}. The horizontal dashed line corresponds to the physically relevant short distance length scale $a=\rho_0^{-1}$; note that $a$ is of order $\rho_0^{-1}$ for all $\gamma$, and for $\gamma \gtrsim 2$, $a \sim \rho_0^{-1}$. The dotted line is a guide for the eye. } 
\label{fig:avg}
 \end{figure}
The agreement between these two independent analyses is seen in \figref{fig:LLFSS}, with full details provided in Appendix~\ref{subsec:analysis}. The two-particle partition EE computed with QMC is also shown in  \figref{fig:LLFSS}, scaled by a factor of $n=2$, demonstrating consistency with the conjectured generic scaling form: 
\begin{equation}
S_2(n) \simeq (n/K)\log N + \cdots
\end{equation}
The small offset between the $n=1$ and $n=2$ data suggests that the constant term in Eq.~\eqref{eq:S2LL} may depend on $n$.

\begin{figure}
\begin{center}
\scalebox{1}{\includegraphics[width=\columnwidth]{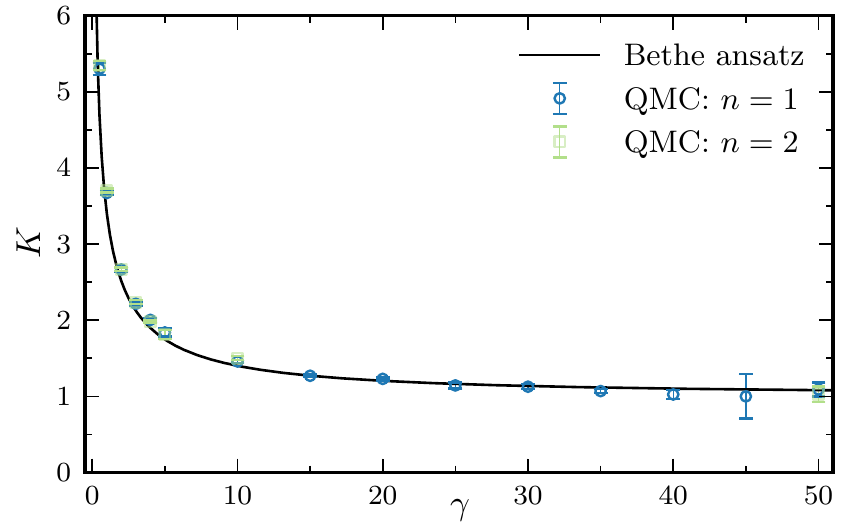}}
\end{center}
\caption{(Color online) The Luttinger parameter ($K$) vs. interaction strength ($\gamma$) for the Lieb-Liniger model determined from the leading order coefficient of the finite-size scaling of $S_2(n)$ as computed by quantum Monte Carlo for $n=1,2$. The line displays the exact value of $K(\gamma)$ known from the Bethe ansatz.} 
\label{fig:LL-Kvgam}
\end{figure}

Using the generic Luttinger liquid three parameter fit to the QMC finite size scaling data shown in \figref{fig:LLFSS}, we have determined the Luttinger parameter from $K=n/b_1$ and plot it as a function of interaction strength $\gamma$ in Fig.~\ref{fig:LL-Kvgam}.  
%
%
The agreement between the QMC derived data points and the analytical value of $K(\gamma)$ determined from the Bethe ansatz provides quantitative confirmation of the predictions from LL theory discussed above.  We reiterate that this analysis uses no microscopic information and is applicable to any bosonic model displaying LL behavior. 

 \section{Discussion} 

We have demonstrated from bosonic Luttinger liquid theory that the Luttinger parameter $K$ may be extracted from the coefficient of the leading logarithmic term in the finite-size scaling of the $n$-particle partition \ren entanglement entropy. We note that this $(n/K) \log N$ scaling is intuitively sensible: this leading term vanishes in the noninteracting limit where $K\rightarrow\infty$ and converges to the particle partition EE of a Mott insulator as $K\rightarrow1$ where the bosons are impenetrable.  Additionally, this result is consistent with previously reported results for the Tonks-Girardeau model of impenetrable bosons, corresponding to $K\rightarrow1$~\cite{Santachiara2007,Haque2009}. It is interesting to contrast the particle partition EE scaling with that of a spatially bipartitioned Luttinger liquid, for which the coefficient of the EE has a universal value of $1/3$ \cite{Vidal2003,Calabrese2004}. As another point of comparison,  bipartite fluctuations in a LL scale logarithmically with subsystem size with a prefactor that is proportional to $K$, rather that $1/K$~\cite{Song2010,Song2011}. Additionally, we have shown that $K$ may be numerically computed via direct quantum Monte Carlo simulations of a microscopic model without appealing to an analysis of the energy or the algebraic decay of correlation functions.  Thus the particle partition entanglement entropy of a bosonic Luttinger liquid provides access to the sole non-universal dimensional parameter which characterizes its long distance behavior.  This suggests that further studies of entanglement entropy under particle partitions offer the possibility of uncovering new insights into quantum phases of matter that are distinct from those employing spatial bipartitions.

 \section{Acknowledgments}
Computations were performed on the Vermont Advanced Computing Core supported by NASA
(NNX-08AO96G).

\appendix

 \section{Path-integral ground state Monte Carlo simulations of the Lieb-Linger model}

Here we describe the quantum Monte Carlo method used to compute the particle partition \ren EE of the Lieb-Liniger model~\cite{Ceperley1995,Hastings2010,Herdman2014a,Herdman2014}.

\subsection{Pair product approximation for the Lieb-Linger model}
\label{app:PPLL}

 $W_{\rm int}$ is determined by the exact two-body propagator for the relative motion:
\begin{equation}
W_{\rm int} \left(x,x',\tau,\lambda\right) \equiv \frac{\rho_{\rm rel}\left(x,x';\tau,2\lambda\right)}{\rho_{\rm 0}\left(x-x';\tau,2\lambda\right)}, \label{eq:Wint}
\end{equation}
where the relative pair propagator $\rho_{\rm rel}$ is defined as
\begin{equation}
\rho_{\rm rel} \left(x,x';\tau,2\lambda \right) = \ebra{x} e^{-\tau H_{\rm CM} } \eket{x'}
\end{equation}
and the center of mass Hamiltonian is 
\begin{equation}
H_{\rm CM} \equiv -2\lambda \frac{\partial^2}{\partial x^2} + V \left(x \right). \label{eq:Hcm}
\end{equation}

For a $\delta$-function potential, the relative propagator is known exactly \cite{Gaveau1986,Lawande1988,Manoukian1989,Cheng1993,Yearsley2008} and can be written in dimensionless form as
\begin{widetext}
\begin{equation}
\tilde{\rho}_{\rm \delta} \left(\tilde{x},\tilde{x}' \right) = \tilde{\rho_0} \left(\tilde{x}-\tilde{x}'\right)-\frac{\ell_0}{2\ell_{\rm int}} \exp\Bigr[\frac{1}{2}\left(\frac{\ell_0}{\ell_{\rm int}}\right)^2+\frac{\ell_0}{\ell_{\rm int}}\Bigl( \left \vert \tilde{x} \right \vert + \left \vert \tilde{x}' \right \vert\Bigr)\Bigl] \mathrm{erfc} \left[ \frac{1}{\sqrt{2}} \left( \frac{\ell_0}{\ell_{\rm int}} +  \left \vert \tilde{x} \right \vert + \left \vert \tilde{x}' \right \vert\right) \right] \label{eq:rhotildedelta2}
\end{equation}
where the two fundamental length scales are defined as $\ell_0 \equiv \sqrt{2\lambda \tau}$ and $\ell_{\rm int} \equiv 2\lambda/g$,
and the associated dimensionless separations are $\tilde{x} \equiv x/\ell_0$. We then may write the interaction weight as
\begin{equation}
W_{\rm int} \left(x,x',\ell_0,\ell_{\rm int} \right) =1-\sqrt{\frac{\pi}{2}}\frac{\ell_0}{\ell_{\rm int}} \exp\Bigr[\frac{1}{2}\left(\frac{\ell_0}{\ell_{\rm int}}\right)^2+\frac{1}{2\lo^2}\left(x-x'\right)^2+\frac{\lo}{\li}\left( \frac{\left \vert x \right \vert}{\lo}+\frac{\left \vert x' \right \vert}{\lo} \right)\Bigl] \mathrm{erfc} \left[ \frac{1}{\sqrt{2}} \left( \frac{\ell_0}{\ell_{\rm int}} + \frac{\left \vert x \right \vert}{\ell_0}+\frac{\left \vert x' \right \vert}{\ell_0}\right) \right].
\end{equation}
\end{widetext}
A similar pair-propagator was used in a quantum Monte Carlo method to study a system of harmonically trapped fermions with $\delta$-function interactions in one dimension in Ref.~\onlinecite{Casula2008}.

\subsection{Convergence of particle partition \ren entropy with QMC parameters}

Here we demonstrate the convergence of $S_2(n)$ to the exact ground state values with imaginary time length $\beta$ and finite time step $\tau$ which parametrize the systematic errors in the QMC method. We use a constant trial wave function in all cases; a variationally optimized trial wave function would be expected to improve convergence to the ground state at smaller $\beta$.

\begin{figure}[t!]
\begin{center}
\includegraphics[width=0.95\columnwidth]{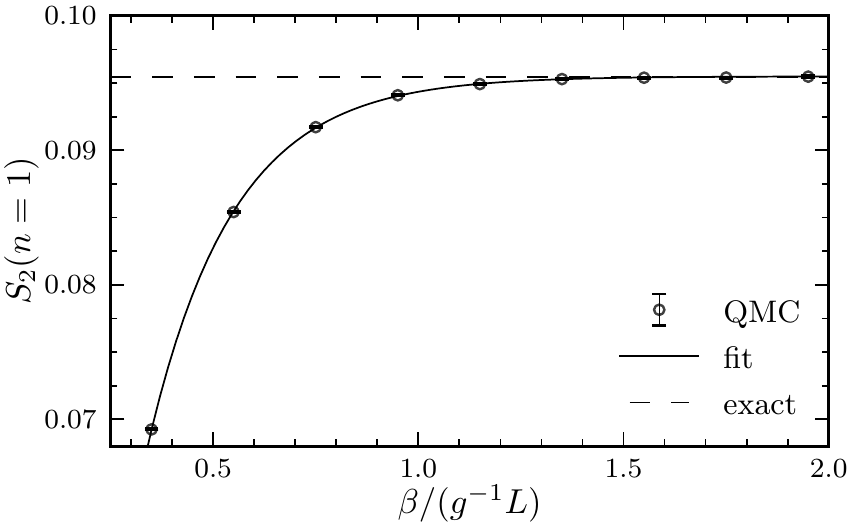}
\includegraphics[width=0.95\columnwidth]{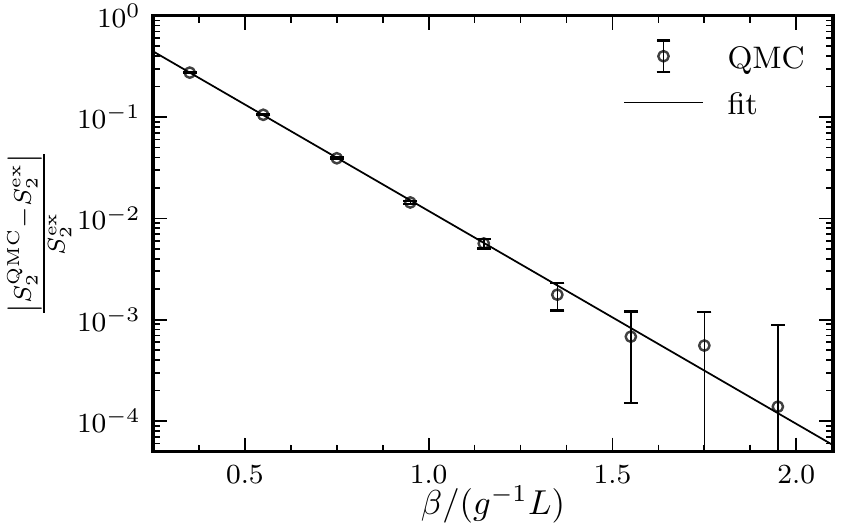}
\end{center}
\caption{ Scaling of $S_2(n=1)$ with imaginary time length $\beta$ for $N=2$ and $\gamma = 5$. (left)  $S_2(n=1)$ as computed by QMC (circles) and the exact value determined by the Bethe ansatz (dashed horizontal line). (right) Fractional error in $S_2(n=1)$ as computed by QMC. The solid lines represent the fit to an exponential decay.} 
\label{fig:S1N2vbeta}
 \end{figure}

To compute ground state properties with imaginary time projection, one must choose a $\beta$ larger than the finite-size gap to excited states. We assume an exponential decay to the ground state value of the form:
\begin{equation}
S \left( \beta \right) = S_0 + c_\beta~e^{-\delta \beta}
\end{equation}
where $c_\beta$ is a dimensionless constant and $\delta$ has units of energy. For $N=2$ we benchmarked our code against the exact ground state value of $S_2(n=1)$ computed from numerical integration of the Bethe ansatz ground state wave function. \figref{fig:S1N2vbeta} shows the convergence of $S_2(n=1)$ with $\beta$ for $N=2$, $\gamma=5$; we find $c_\beta = -0.142(1)$, $\delta = 4.82(2) gL^{-1}$, and $S_0 = 0.09549(2)$ which is in agreement with the exact ground state value of $S_2^{\rm{BA}} = 0.09546$ down to order $10^{-5}$.
%
%
%
\begin{figure}[t]
\begin{center}
\includegraphics[width=0.95\columnwidth]{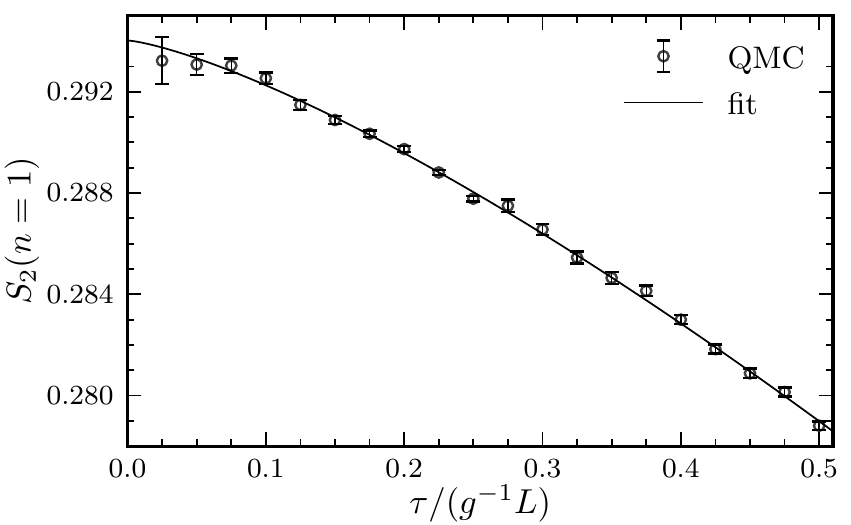}
\includegraphics[width=0.95\columnwidth]{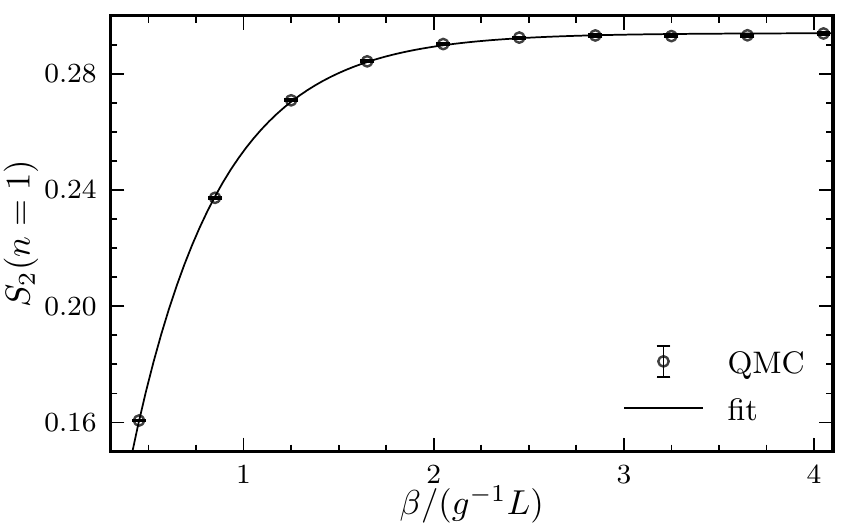}
\end{center}
\caption{Scaling of the $S_2(n=1)$ with (left) imaginary time step, $\tau$, and (right) imaginary time length, $\beta$, for  $N=4$, $\gamma=5$. The solid lines represent the best fits to (left) power-law and (right) exponential scaling.} 
\label{fig:S1N4}
 \end{figure}

\figref{fig:S1N4} shows the imaginary-time length scaling for $N=4$, $\gamma=5$ for which we find and exponential decay with $c_\beta=-0.354(3)$, $\delta=2.16(2) gL^{-1}$, and $S_0=0.2940(3)$. When scaling $N$ at constant $\gamma$, we take $\beta \sim N$ due to the $1/L$ scaling of the finite-size Luttinger liquid gap.


We assume a power-law scaling of the systematic error due to finite time-step $\tau$, for fixed $g$ and $L$, of the form
\begin{equation}
S \left( \tau \right) = S_0 + c_\tau~\left(\frac{\tau}{g^{-1}L}\right)^\nu
\end{equation}
where $c_\tau$ and $\nu$ are dimensionless constants (that may in general depend on $g$ and $L$).
\figref{fig:S1N4} shows the convergence of $S_2(n=1)$ with $\tau$ for $N=4$, $\gamma=5$ for which we find $c_\tau=-0.038(1)$, $\nu = 1.33(5)$, and $S_0 = 0.2940(3)$. The required $\tau$ for a particular systematic error scales primarily as a function of interaction and density and therefore $\gamma$ and thus is independent of system size.

\subsection{Analysis of QMC data}
\label{subsec:analysis}
\begin{figure*}[t]
\begin{center}
\subfloat{\scalebox{1}{\includegraphics[width=\columnwidth]{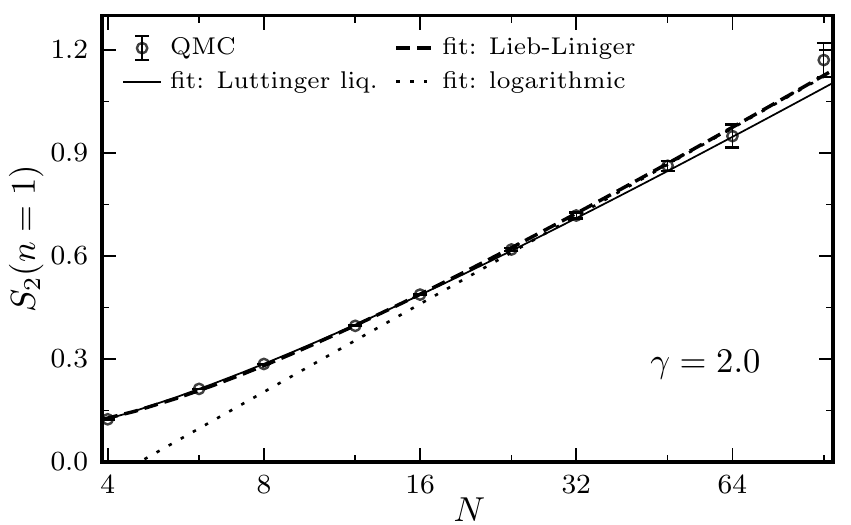}}}
\subfloat{\scalebox{1}{\includegraphics[width=\columnwidth]{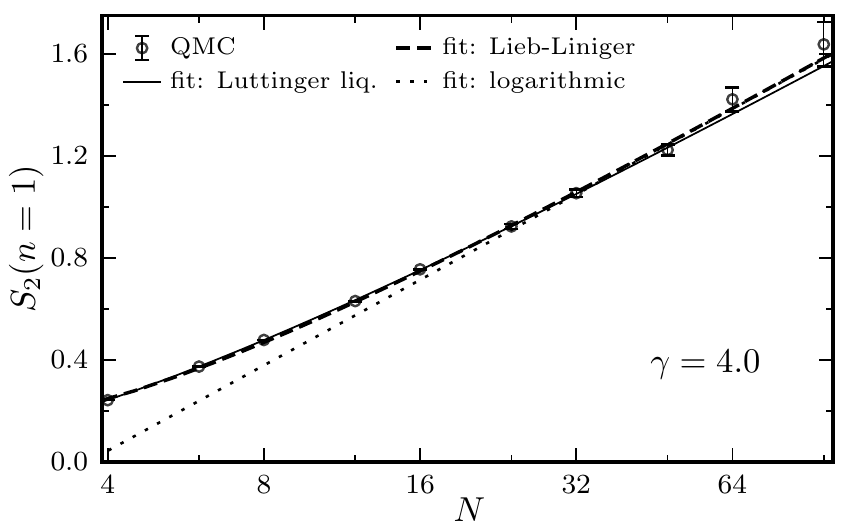}}}  
\\
\vspace{-1.1\baselineskip}
\subfloat{\scalebox{1}{\includegraphics[width=\columnwidth]{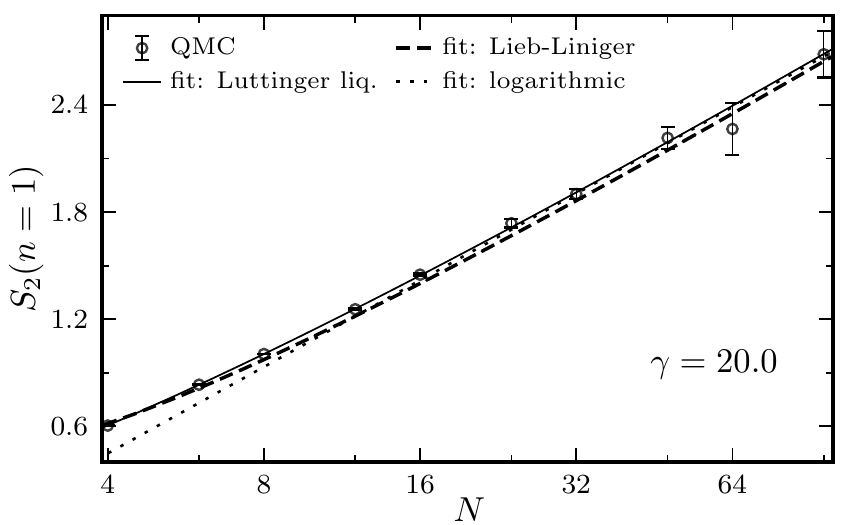}}}
\subfloat{\scalebox{1}{\includegraphics[width=\columnwidth]{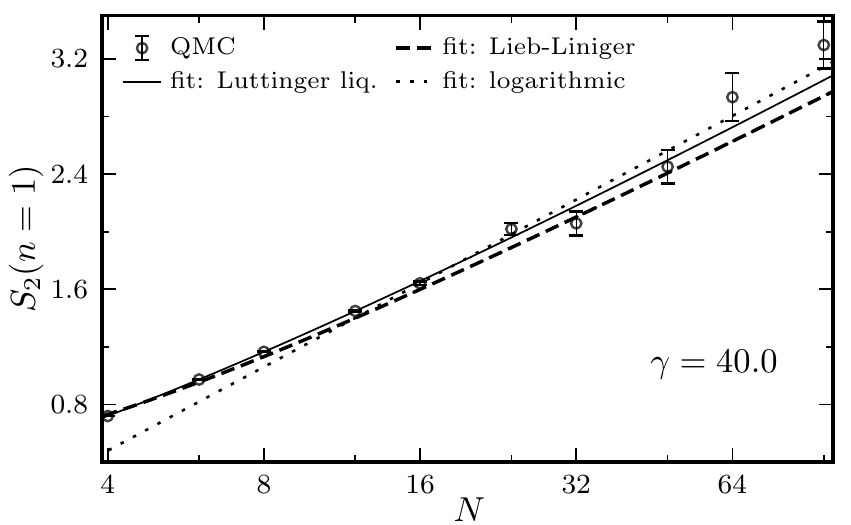}}}
\end{center}
\caption[justification=raggedright]{
 Finite-size scaling of $S_2(n=1)$ for $\gamma\in\{2,4,20,40\}$ as computed by QMC. The lines represent fits to the generic Luttinger liquid scaling form (solid line), finite-size Lieb-Liniger form (dashed line) and pure logarithmic scaling with no finite-size corrections (dotted line).
}
\label{fig:SvsNb}
\end{figure*}

We now discuss in detail the analysis of the finite-size scaling of $S_2(n)$ as computed by QMC. We consider three scaling functions: the generic Luttinger liquid form given in Eq.~\eqref{eq:S2NFS}, the finite-size Lieb-Linger specific form using Eq.~\eqref{eq:S2NFS1}, and a pure logarithmic form, valid only at large $N$.
\begin{itemize}
\item {\it Luttinger liquid:}
In fitting to the generic Luttinger liquid form given in Eq.~\eqref{eq:S2NFS}, we take all three parameters $\{b_1,b_2,b_3\}$ to be free parameters determined numerically by the fit; this approach uses no assumptions about the microscopic model. Thus this fit allows us to extract $K$ numerically from the finite-size scaling data by taking $K=n/b_1$. 
\item {\it Finite-size Lieb-Liniger:}
    The finite-size Lieb-Linger specific form given in Eq.~\eqref{eq:S2NFS1} is employed, where the short distance behavior of the one-body RDM $\Phi(r/a)$ is taken to be that of the Lieb-Linger model given in \eqref{eq:LLfr}. In this case, for each interaction strength $\gamma$ we fix $K$, $c_2$ and $c_3$ to take their known Bethe ansatz values, leaving only a single free parameter, the short distance cutoff $a$. This approach allows us to test the Lieb-Linger predicted form against the QMC data.
\item {\it Pure logarithmic:}
We also consider a generic two parameter fit to a simple logarithmic scaling in $N$, {\it i.e.} taking $b_3 = 0$ in \eqref{eq:S2NFS}. This again allows us to numerically extract the $K$ from $b_1$ with no assumptions about the microscopic model. 
\end{itemize}
Fits to the finite-size scaling QMC data for $\gamma \in \{2,4,20,40\}$ using all three of these forms are shown in \figref{fig:SvsNb}. 
%
%
For the pure logarithmic case, we have only fit to the larger system sizes as there are clear deviations from the pure logarithmic scaling for small $N$. In the other approaches we fit to all values of $N$ for $4 \leq N \leq 96$. To extract $K$ numerically with the same precision shown in \figref{fig:LL-Kvgam} (which uses the full generic Luttinger liquid scaling form where $b_3$ is not set to zero) with a pure logarithmic fit ({\it i.e.} $b_3=0$) would require studying larger system sizes due to the finite-size corrections that appear for small $N$. 

\bibliographystyle{apsrev4-1}
\bibliography{PartEEinLL}

\end{document}